\documentclass[11pt,english]{article}
\pdfoutput=1
\usepackage{lmodern}
\usepackage[T1]{fontenc}
\usepackage[latin9]{inputenc}
\usepackage{geometry}
\usepackage{array}
\usepackage{graphicx}
\usepackage{setspace}
\usepackage{esint}
\makeatletter

\providecommand{\tabularnewline}{\\}

\newcommand{\lyxaddress}[1]{
\par {\raggedright #1
\vspace{1.4em}
\noindent\par}
}

\date{}

\makeatother

\usepackage{babel}
\begin{document}

\title{Charged scalar gravitational collapse in de Sitter spacetime}

\author{Cheng-Yong Zhang$^{1}$\thanks{zhangcy@sjtu.edu.cn}, Shao-Jun Zhang$^{2}$\thanks{sjzhang84@hotmail.com },
De-Cheng Zou$^{3}$\thanks{dczou@yzu.edu.cn}, Bin Wang$^{1}$\thanks{wang\_b@sjtu.edu.cn}}

\maketitle

\lyxaddress{\begin{center}
\textit{1. Center of Astronomy and Astrophysics, Department of Physics
and Astronomy,}\\
\textit{Shanghai Jiao Tong University, Shanghai 200240, China}\\
\textit{2. Institute for Advanced Physics and Mathematics,
Zhejiang University of Technology, Hangzhou, 310032, China
}\\
\textit{3. College of Physics Science and Technology, Yangzhou University,
Jiangsu 225009, China}\\

\par\end{center}}
\begin{abstract}
We study the charged scalar collapse in de Sitter spacetimes.  With the electric charge, there is one more competitor to join the competition of dynamics in the gravitational collapse. We find that  two factors can influence the electric charge. If we just  adjust the charge conjugation, the electric charge effect is always perturbative at the black hole threshold.  The electric charge can also be influenced by the initial conditions of perturbations.  These initial parameters can be tuned to control the competition in dynamics and present us  new and rich physics in the process of gravitational collapse.   We give physical explanations of these phenomena found in dynamics. Furthermore we show that the properties of the gravitational collapse we observed do not depend on spacetime dimensions.

\end{abstract}
PACS number: 04.70.-s, 04.25.Dm

\section{Introduction}

Gravitational collapse is one of the most intriguing phenomena in
the theory of gravity. In the past two decades, a lot of interesting phenomenologies have been found at the threshold of the black hole formation in gravitational collapse models. The first discovery of the behavior of the gravitational collapse was found by Choptuik \cite{Choptuik-93} who studied the self-gravitating dynamics of massless scalar field in a spherically symmetric background. Although this is the first and simplest realistic model of the gravitational collapse, it contains most of the features associated with critical collapse.  There exists a one-parameter family of initial data in the collapse of the spherically symmetric massless scalar field;  if the parameter value $p$ is over the critical value $p_{\star}$, a black hole will be formed; however if  $p<p_{\star}$, the scalar field eventually dissipates away. The black hole mass has been found with a power-law scaling, $M\sim(p-p_{\star})^{\gamma}$, where the critical exponent $\gamma$ is observed to be universal. For more general discussions and understandings of the gravitational collapse, please refer to review articles \cite{AnZhong-01,Gundlach-07} and references therein.

The physical reason behind the gravitational collapse is the nature of competition in the dynamics \cite{Choptuik-98}. The kinetic energy of the imploding scalar field wants to disperse the field to infinity, however the gravitational potential energy, if it is sufficiently dominant during the collapse,  will trap some amount of energy of the system to form a black hole. The interesting key point is that the dynamical competition can be controlled by tuning a parameter in the initial conditions.  If $p$ in the one-parameter family of initial data is less than some critical value  $p_{\star}$, the scalar field will completely run away. But if $p>p_{\star}$, a black hole can be developed.

Besides the internal factors in the initial conditions of perturbations, for example the initial field amplitude etc, which control the dynamics of the gravitational collapse, it is interesting to ask whether some external factors can influence the perturbations, affect the imploding of the scalar field and finally control the gravitational collapse. One of the external factors we can think about is the cosmological constant. Most of the critical phenomena in the gravitational collapse were studied in the asymptotically flat spacetime. It is of interest to investigate how the spacetimes with cosmological constant influence the gravitational collapse. Recently in the anti-de Sitter (AdS) background with negative cosmological constant, the gravitational collapse was examined in the spherically symmetric configuration. It was argued that in the Einstein gravity the AdS space is unstable and the AdS black hole can be formed under arbitrarily small generic perturbations \cite{Bizon-11}. This instability is due to the subtle interplay of local nonlinear dynamics and the nonlocal kinematical effect of the AdS reflecting boundary. It was further claimed that the results of the AdS collapse of a scalar field exist as well in other dimensions \cite{Jalmuzna-11}. This clarified the argument in \cite{Garfinkle-11} that for dimensions higher than four there is no black hole formation in AdS due to the small initial perturbation of the scalar field. Considering the dependence of the instability and turbulent behavior on the local dynamics, recently the stability of AdS spacetime has been examined further in the Einstein-Gauss-Bonnet gravity \cite{Deppe-14}. It was found that with the Gauss-Bonnet factor, the AdS black hole cannot be formed dynamically if the total mass/energy content of the spacetime is too small. In the AdS spacetime, the collapse of self-interacting scalar field was examined in \cite{Cai-15}. The gravitational collapses in the de Sitter spacetime with an addition of a positive cosmological constant have also been studied. In \cite{Hod-96}, it was claimed that the positive cosmological constant lead to a dissipation of the critical solution. In this work, we will examine this effect more carefully in the gravitational collapse in different dimensional de Sitter spacetimes. Another study of the gravitational collapse in the de Sitter spacetime was done in \cite{Iwashita-05}, where the attention was on discussing whether the gravitational collapse disturbs the dS/CFT correspondence.

Another external factor that can influence the gravitational collapse in the spherically symmetric spacetime is the electric charge. Including the electric charge, we will have one more repulsive force to join the competition in dynamics. Intuitively this external factor will hinder the gravitational collapse so that we need a stronger amplitude in the initial perturbation to develop a black hole. A simple model for describing the charged matter collapse is to consider a complex scalar field coupled to electromagnetism. In \cite{Hod-96}, the influence of the charge conjugation on the formation of a charged black hole was studied. Surprisingly it was observed that with the external factor (i.e. the charge conjugation), the black hole can be formed
even in the former subcritical region. For the supercritical initial condition with stronger initial perturbation,  the formed black hole can have larger mass due to the charge conjugation. It was explained that although the charge opposes gravitation, the electromagnetic energy density also contributes to the gravitational binding and by doing so the formation of a black hole becomes easier due to this external factor. This result was confirmed in the study of the scalar electrodynamics \cite{Gundlach-96}.  Furthermore the authors also obtained the universal power-law scaling of the black hole charge.  Some related works about charged scalar collapse in several setups can be found in \cite{Nakonieczna-11,Nakonieczna-12,Nakonieczna-15}. In this work, we  further examine the effect of the electric charge and try to answer the question of whether the black hole charge is always a perturbation and does not play a role in the competition of dynamics in the process of collapse. We will concentrate more on the evolution of the charged scalar perturbation and examine how  the process of the gravitational collapse of the charged scalar field and the time scale of the black hole formation can be affected by initial conditions and the electric charge.

In the study of the stability of the Reissner-Nordstrom-dS
black holes, it was found that the black hole is gravitationally unstable for large values of
the electric charge and cosmological constant when the spacetime dimension
$D\geq 7$ \cite{Konoplya-08,Cardoso-10}. These results were obtained by numerical study on the quasinormal modes in the linear perturbation level \cite{Konoplya-08} and later the instability was confirmed analytically \cite{Cardoso-10}. However deep physical reasons behind this phenomenon are still lacking. This result leads us to raise the question of whether the behaviors of the gravitational  collapse of charged scalar field  in de Sitter background are different in different spacetime dimensions. In asymptotically flat spacetime background, scalar field collapses in higher dimensional spacetime were found similar to four dimensions \cite{Garfinkle-99}. Here we will examine whether the external factors, the cosmological constant and the electric charge can modify the result. Instead of the linear perturbation, our investigation will disclose the nature and consequences of nonlinearity in general relativity. In our analysis we will concentrate on spacetime dimension $D=7$, which reflects the special property in the stability of the highly charged de Sitter black hole, and later we will compare the result obtained in $D=7$ with other spacetime dimensions for the gravitational collapse of the charged scalar field in the de Sitter background.

This paper is organized as follows. In section 2 we will derive the evolution
equations in spherically symmetric dS spacetime. In section 3, we will describe
the numerical method and introduce the initial conditions for the evolution of perturbations.
In section 4 we will show numerical results. The conclusion will be provided in
the last section.

\section{The evolution equations}
We consider a spherically symmetric charged scalar field $\phi$, which is complex in combination of two real scalar fields $\phi_1, \phi_2$ in the form
$\phi=\phi_{1}+i\phi_{2}$. The electromagnetic field tensor is defined as $F_{ab}=2A_{[b;a]}$.  The total Lagrangian of the scalar field and the electromagnetic field is
\begin{eqnarray}
\mathcal{L} & = & -\frac{1}{2}g^{ab}\left(\partial_{a}\phi+iqA_{a}\phi\right)\left(\partial_{b}\phi^{*}-iqA_{b}\phi^{*}\right)-\frac{1}{16\pi}F_{ab}F_{cd}g^{ac}g^{bd},
\end{eqnarray}
where $q$ is the coupling between the scalar field and the electromagnetic field. Here
 $\phi^{*}$ is the complex conjugate of $\phi$.

Varying $\mathcal{L}$
with respect to $\phi^{*}$ and $A_{a}$ respectively, one gets the equation of
motion of scalar field
\begin{eqnarray}
\nabla^{a}\nabla_{a}\phi+iqA^{a}\left(2\partial_{a}\phi+iqA_{a}\phi\right)+iq\nabla_{a}A^{a}\phi & = & 0,\label{eq:scalarEq}
\end{eqnarray}
and the equation of motion of the electromagnetic field
\begin{eqnarray}
\frac{1}{4\pi}\nabla_{b}F_{a}^{\ b}-iq\phi\left(\partial_{a}\phi^{*}-iqA_{a}\phi^{*}\right)+iq\phi^{*}\left(\partial_{a}\phi+iqA_{a}\phi\right) & = & 0.\label{eq:MaxwellEq}
\end{eqnarray}

The conserved current is described by
\begin{equation}
J_{a}=-8\pi q\left[\phi_{2}\partial_{a}\phi_{1}-\phi_{1}\partial_{a}\phi_{2}-qA_{a}\left(\phi_{2}^{2}+\phi_{1}^{2}\right)\right],
\end{equation}
which satisfies $\nabla_{a}J^{a}=0$. The corresponding charge can be obtained from the integration
\begin{eqnarray}
Q & = & -\int d^{D-1}x\sqrt{\gamma}n_{a}J^{a},\label{eq:Charge}
\end{eqnarray}
where $n_{a}$ is the normal vector of spacelike hypersurface,
$\gamma$ the determinant of the induced metric and $Q$ the total
charge surrounded by this hypersurface.

The Einstein equation reads
\begin{eqnarray}
G_{ab}+\Lambda g_{ab} & = & 8\pi T_{ab},
\end{eqnarray}
where $\Lambda$ is the cosmological constant which is positive in the de Sitter spacetime and the energy-momentum tensor is described by
\begin{eqnarray}
T_{ab} & = & \frac{1}{2}\left(\partial_{a}\phi\partial_{b}\phi^{*}+\partial_{a}\phi^{*}\partial_{b}\phi\right)+\frac{1}{4\pi}F_{ac}F_{bd}g^{cd}+q^{2}A_{a}A_{b}\phi^{*}\phi+\mathcal{L}g_{ab}\nonumber \\
 &  & +\frac{1}{2}\left(-iqA_{b}\phi^{*}\partial_{a}\phi+iqA_{b}\phi\partial_{a}\phi^{*}-iqA_{a}\phi^{*}\partial_{b}\phi+iqA_{a}\phi\partial_{b}\phi^{*}\right).
\end{eqnarray}

We restrict to the spherically symmetric spacetime with  the metric
\begin{eqnarray}
ds^{2} & = & -a(t,r)e^{-2\delta(t,r)}dt^{2}+\frac{1}{a(t,r)}dr^{2}+r^{2}d\Omega_{D-2}^{2}.
\end{eqnarray}
 where $d\Omega_{D-2}^{2}$ is the line element of the unit $(D-2)$-dimensional
sphere. The radial coordinate $r$ measures the proper surface area.
The normal vector of the spacelike hypersurface $t=const$ then is
$n_{a}=\sqrt{a}e^{-\delta}(dt)_{a}$. In the spherically symmetric spacetime,
the gauge field can be written as
\begin{equation}
A_{\mu}=(A_{t}(t,r),A_{r}(t,r),0,0,0,....).
\end{equation}
 We take the axial gauge and set further $A_{r}=0$, the only non-vanishing
electromagnetic field is then $F_{tr}=-\partial_{r}A_{t}$.

Introducing auxiliary variables of the scalar field
\begin{eqnarray}
\Pi_{1}(t,r)\equiv a^{-1}e^{\delta}\left(\partial_{t}\phi_{1}-qA_{t}\phi_{2}\right) & , & \Phi_{1}(t,r)\equiv\partial_{r}\phi_{1},\label{eq:AuxiliaryField}\\
\Pi_{2}(t,r)\equiv a^{-1}e^{\delta}\left(\partial_{t}\phi_{2}+qA_{t}\phi_{1}\right) & , & \Phi_{2}(t,r)\equiv\partial_{r}\phi_{2},\nonumber
\end{eqnarray}
 the scalar equation (\ref{eq:scalarEq}) turns into
\begin{eqnarray}
\partial_{t}\Phi_{1} & = & \partial_{r}\left(ae^{-\delta}\Pi_{1}+qA_{t}\phi_{2}\right),\nonumber \\
\partial_{t}\Phi_{2} & = & \partial_{r}\left(ae^{-\delta}\Pi_{2}-qA_{t}\phi_{1}\right),\nonumber \\
\partial_{t}\Pi_{2} & = & r^{2-D}\partial_{r}\left(ae^{-\delta}r^{D-2}\Phi_{2}\right)-qA_{t}\Pi_{1},\label{eq:scalar2timeder}\\
\partial_{t}\Pi_{1} & = & r^{2-D}\partial_{r}\left(ae^{-\delta}r^{D-2}\Phi_{1}\right)+qA_{t}\Pi_{2}.\nonumber
\end{eqnarray}

Introducing an auxiliary variable for the gauge field
\begin{eqnarray}
E(t,r) & = & -e^{\delta}\partial_{r}A_{t},\label{eq:E2}
\end{eqnarray}
we have the evolution equation of the electromagnetic field (\ref{eq:MaxwellEq}) in the form
\begin{eqnarray}
\partial_{t}E & = & -8\pi qae^{-\delta}\left(\phi_{2}\Phi_{1}-\phi_{1}\Phi_{2}\right).\label{eq:Maxwell2}
\end{eqnarray}

 The  constraint equation between the scalar and the electromagnetic fields has the form
\begin{equation}
r^{2-D}\partial_{r}\left(r^{D-2}E\right)+8\pi q\left(\phi_{2}\Pi_{1}-\phi_{1}\Pi_{2}\right)=0.\label{eq:MaxwellConstraint}
\end{equation}

The nontrivial Einstein equations expressed in
terms of auxiliary variables are
\begin{eqnarray}
\partial_{r}\delta & = & -\frac{8\pi r}{(D-2)}\left(\Pi_{1}\Pi_{1}+\Pi_{2}\Pi_{2}+\Phi_{1}\Phi_{1}+\Phi_{2}\Phi_{2}\right),\label{eq:delta2}\\
\partial_{r}a & = & \frac{(D-3)(1-a)}{r}-\frac{2r}{(D-2)}E^{2}-\frac{2r}{(D-2)}\Lambda\label{eq:a2}\\
 &  & -\frac{8\pi ra}{(D-2)}\left(\Pi_{1}\Pi_{1}+\Pi_{2}\Pi_{2}+\Phi_{1}\Phi_{1}+\Phi_{2}\Phi_{2}\right).\nonumber
\end{eqnarray}
Furthermore, there is also a momentum constraint requirement
\begin{equation}
\partial_{t}a=-\frac{16\pi r}{(D-2)}a^{2}e^{-\delta}\left(\Pi_{1}\Phi_{1}+\Pi_{2}\Phi_{2}\right).\label{eq:momentum-constr}
\end{equation}

In the spherically symmetric spacetime, the conserved charge (\ref{eq:Charge})
can be expressed into
\begin{eqnarray}
Q(t,r) & = & -16q\frac{\pi^{\frac{D+1}{2}}}{\Gamma(\frac{D-1}{2})}\int_{0}^{r}r^{D-2}\left(\phi_{2}\Pi_{1}-\phi_{1}\Pi_{2}\right)dr,\label{eq:Charge-1}
\end{eqnarray}
in which we have used the volume of $(D-2)$-dimensional sphere $V_{D-2}=\frac{2\pi^{\frac{D-1}{2}}}{\Gamma(\frac{D-1}{2})}$. It is easy to see that besides the charge conjugation $q$ between the scalar field and the electromagnetic field, the amplitudes of $\phi_i$ and $\Pi_i$ also influence the value of the charge. In \cite{Hod-96}, the amplitudes of  $\phi_i$ and $\Pi_i$ were fixed to be the same and only the influence of the value of the charge conjugation q on the perturbation and the gravitational collapse was discussed. In the following we will confirm the result obtained in \cite{Hod-96} that the charge conjugation effect on the electric charge can always be treated perturbatively at the black hole threshold. Furthermore we will disclose more properties of the electric charge on the gravitational collapse provided that we set free the amplitudes of initial perturbations.

Moreover we define the  mass aspect function $M(t,r)$  in analogy with the
usual Reissner-Nordstrom--dS form of the static spherically symmetric metric
\begin{eqnarray}
M(t,r) & = & \frac{r^{D-3}}{2}\left[1-a(t,r)+\frac{2Q(t,r)^{2}}{(D-3)(D-2)r^{2(D-3)}}-\frac{2\Lambda}{(D-1)(D-2)}r^{2}\right],\label{eq:Mass}
\end{eqnarray}
 and the gravitational potential is in the form $V(t,r)=\frac{1-g{}_{00}(t,r)}{2}=\frac{1}{2}\left(1+a(t,r)e^{-2\delta(t,r)}\right)$.

\section{Numerical method}

To solve the evolution equations, we need to specify initial conditions.
Since the equations of motion of charged scalar field (11) are complex second-order
differential equations, four initial profiles are needed to be specified.
We take the following initial profiles:
\begin{eqnarray}
\phi_{1}(0,r)=a_{1}e^{-\left(\frac{r-r_{1}}{\sigma_{1}}\right)^{2}} & , & \phi_{2}(0,r)=a_{2}e^{-\left(\frac{r-r_{2}}{\sigma_{2}}\right)^{2}},\nonumber \\
\Pi_{1}(0,r)=a_{3}e^{-\left(\frac{r-r_{3}}{\sigma_{3}}\right)^{2}} & , & \Pi_{2}(0,r)=a_{4}e^{-\left(\frac{r-r_{4}}{\sigma_{4}}\right)^{2}},\label{eq:initial2}
\end{eqnarray}
where the initial amplitudes are taken differently.  Given initial conditions (\ref{eq:initial2}), the auxiliary electromagnetic
field $E$ can be determined from (\ref{eq:MaxwellConstraint}) and the
initial metric coefficients can be calculated from (\ref{eq:delta2}) and (\ref{eq:a2}). Then with the help of  (\ref{eq:E2}), we can derive the initial gauge
field $A_{t}$. Until this moment the
initial conditions can totally be determined. Furthermore from (\ref{eq:scalar2timeder}), we know how the charged scalar field evolves, then from (\ref{eq:Maxwell2}) we can learn how the electromagnetic field evolves, and using (\ref{eq:delta2}) and (\ref{eq:a2}) we know how the metric coefficients change. Later with (\ref{eq:E2}) and (\ref{eq:AuxiliaryField}) we know how the gauge field and scalar field develop. Repeating this procedure, we can obtain the evolution of the charged scalar field  and examine whether the perturbation can lead to the instability in the spherically symmetric dS spacetime. There are two constraint equations (\ref{eq:MaxwellConstraint}) and (\ref{eq:momentum-constr}) which we have not used, but they are important to check the accuracy of the numerical computations.

Besides the initial conditions, we also need to specify the boundary
conditions in the numerical computations. The regularity of (\ref{eq:a2}) at center $r=0$ requires
that $a(t,0)=1$. We solve the constraints by integrating outwards
and take the boundary condition $\delta(t,0)=0$. This implies that
the time coordinate at the center $r=0$ is chosen as the proper time.
The remaining gauge allows us to take $A_{t}(t,0)=0$.

We solve the system of equations numerically using a fourth-order Runge-Kutta method
in both time and space directions. The adjustable time step $\Delta t$
is kept $1/10<e^{-\delta(R_{c})}\Delta t/\Delta r<1/3$ for the constant spatial
grid spacing $\Delta r$. Here $R_{c}$ is the cosmological horizon developed in the existence of the scalar field and the electromagnetic field, which is determined by the largest root of $a(t,r)=0$. This scheme allows for stable long-time
evolution. To overcome the error caused by $1/r$ in the evolution
equations near $r=0$, we expand the variables around the center.
Smoothness at the origin then implies that we have following power
series expansions near $r=0$:
\begin{eqnarray}
a(t,r) & = & 1+a_{2}(t)r^{2}+a_{4}(t)r^{4}+a_{6}(t)r^{6}+O(r^{7}),\nonumber \\
\delta(t,r) & = & \delta_{2}(t)r^{2}+\delta_{4}(t)r^{4}+\delta_{6}(t)r^{6}+O(r^{7}),\nonumber \\
\phi_{1}(t,r) & = & \phi_{10}(t)+\phi_{12}(t)r^{2}+\phi_{14}(t)r^{4}+O(r^{5}),\\
\phi_{2}(t,r) & = & \phi_{20}(t)+\phi_{22}(t)r^{2}+\phi_{24}(t)r^{4}+O(r^{5}),\nonumber \\
A_{t}(t,r) & = & A_{4}(t)r^{4}+O(r^{5}).\nonumber
\end{eqnarray}
These expansions allow us to express values of functions near the boundary $r=0$
with their values inside the domain. For example if function $f(r)$
has the expansion near $r=0$: $f(r)=f_{0}+f_{2}r^{2}+f_{4}r^{4}+O(r^{5}),$ we get the folllowing equations
\begin{eqnarray}
f(\triangle r) & = & f_{0}+f_{2}(\triangle r)^{2}+f_{4}(\triangle r)^{4},\nonumber \\
f(2\triangle r) & = & f_{0}+f_{2}(2\triangle r)^{2}+f_{4}(2\triangle r)^{4},\\
f(3\triangle r) & = & f_{0}+f_{2}(3\triangle r)^{2}+f_{4}(3\triangle r)^{4},\nonumber
\end{eqnarray}
where $\triangle r$ is the spatial
grid spacing which is very small. Then we can express $f_{0}$ in terms of $f(\triangle r),f(2\triangle r)$ and $f(3\triangle r)$
as
\begin{equation}
f(0)=f_{0}=\frac{15f(\triangle r)-6f(2\triangle r)+f(3\triangle r)}{10}.
\end{equation}
Similarly, we can also write $f(\triangle r)$ in terms of $f(2\triangle r),f(3\triangle r),f(4\triangle r)$.
In this way, the numerical error caused by $1/r$ term is suppressed. Besides, de l'Hopital rule $f/r=f'(r)-r(f/r)'$ is applied at $r=0$
if $f(r)\rightarrow0$ as $r\rightarrow0$. The instability of $f/r$
near $r=0$ is suppressed by the factor $r$ comparing to the leading
term $f'(r)$ \cite{Maliborski-13}. The Kreiss-Oliger dissipation
which is crucial to stabilize the solutions is employed \cite{Kreiss-72}.

In our numerical calculations, the criteria for the formation of a black hole is set as
$a(r_{B})<0.1$, where $r_{B}$ is the apparent horizon of the black hole.
The cosmological constant $\Lambda=\frac{(D-1)(D-2)}{2}$ unless otherwise specified, where $D$ is the spacetime dimension,
so the cosmological horizon of a pure de Sitter spacetime locates
at $r=1$. In all numerical calculations, we take initial conditions
(\ref{eq:initial2}) and choose $a_{2}=0.1,a_{3}=0$. The centers of the initial pulses are  $r_{1}=r_{2}=r_{3}=r_{4}=0$,
the widths $\sigma_{1}=\sigma_{2}=\frac{R_{c}}{10}$, $\sigma_{3}=\sigma_{4}=\frac{R_{c}}{3}$.

\section{Numerical results}

\begin{figure}
\begin{centering}
{\small{}}%
\begin{tabular}{cc}
{\small{}\includegraphics[scale=0.5]{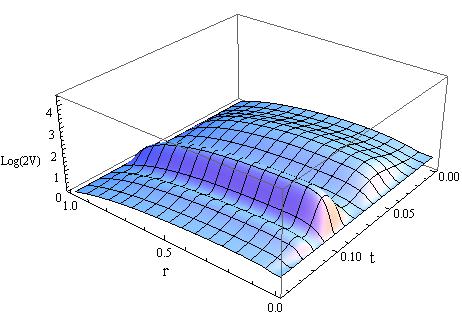}} & \includegraphics[scale=0.5]{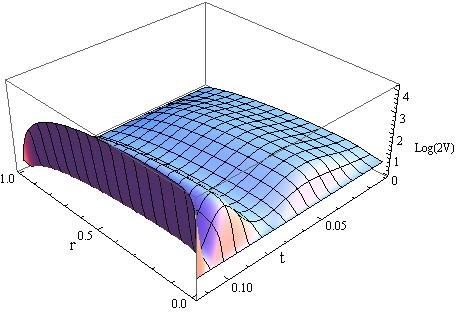}\tabularnewline
\end{tabular}
\par\end{centering}{\small \par}

{\footnotesize{}\caption{{\footnotesize{}\label{fig:potential_1}The gravitational potential behaviors in
the spherically symmetric de Sitter spacetime. The vertical axis is the logarithm of the potential. In the left panel, we start with a subcritical initial condition of the perturbation and the gravitational potential cannot grow to trap the perturbation and disappears finally. In the right panel, we start with a supercritical initial condition of the perturbation and we observe that the gravitational potential can grow to be very high to trap the perturbation field, which results in the formation of a black hole in the gravitational collapse. }}
}{\footnotesize \par}
\end{figure}

We first present a qualitative picture on how gravitational collapse in de Sitter
spacetime can happen. We plot the evolution of the gravitational potential
in fig.(\ref{fig:potential_1}). On the left panel, we present the evolution of the gravitational potential when the initial amplitude of the perturbation is below the critical value, where we observe that the potential barrier
arises at first and then disappears after certain period of time evolution. No high enough gravitational potential can grow to trap the perturbation and the scalar field finally escapes out of the cosmological
horizon. On the right panel, we exhibit the situation when the perturbation field starts with strong initial amplitude and a very high potential barrier can be developed near the center in the evolution so that a big part of the scalar field can be trapped and the black hole can be formed.

{\small{}}
{\small \par}

In the following, besides the initial conditions in the perturbation which are considered as internal factors to influence the gravitational collapse in the de Sitter spacetime, we will examine the collapse of the perturbation under some external factors, such as the cosmological constant and the electric charge. We will also study the effect of the spacetime dimensions. For the addition of the electric charge, considering the interesting stability property of the Reissner-Nordstrom-dS black hole which is special in seven dimensions, we will first illustrate the behavior of the gravitational collapse in seven-dimensional de Sitter space and then report the results in other dimensions.

\subsection{The effect of the positive cosmological constant}

In four-dimensional spherically symmetric de Sitter spaetime, the gravitational collapse was examined in \cite{Hod-96}, where the electric charged was neglected.   It was argued that the positive cosmological constant leads to a dissipation of the perturbation.  This tells us that the cosmological constant, which is an external factor, plays the role similar to the  kinetic energy in the dynamics competition and hinders the process of the gravitational collapse. It is of interest to check whether this property is universal in higher dimensions. Remember that in the asymptotically flat spacetime, scalar field collapse in higher dimensions are similar to four-dimensions \cite{Garfinkle-99}. How about the effect of the cosmological constant in the collapse of scalar perturbation in higher dimensional spherically symmetric de Sitter spacetime?

In order to answer this question, we have carefully calculated the critical initial amplitude value of the perturbation. We specify $\Lambda=0.001$ in this subsection for the convenience of the comparison of the gravitational collapse in dS spacetime and that in asymptotically flat spacetime and to grasp the effect of positive cosmological constant. Without the electric charge, we turn off the imaginary part in the scalar field and can set $\Pi_{1}(0,r)=0 (a_3=0)$ and a Gaussian initial pulse for $\phi_{1}(0,r)$ like the form in (20). We associate the parameter $p$ with the amplitude of the Gaussian, which is $a_1$ in (\ref{eq:initial2}). We will find the critical $p_{\star}$, where black hole formation first occurs in de Sitter spacetime and asymptotically flat spacetime in different spacetime dimensions, respectively. The results are listed in Table \ref{tab:L}.

{\small{}}
\begin{table}
\begin{onehalfspace}
\begin{centering}
{\footnotesize{}}%
\begin{tabular}{|>{\centering}p{0.1\textwidth}|>{\centering}p{0.12\textwidth}|>{\centering}p{0.12\textwidth}|>{\centering}p{0.12\textwidth}|>{\centering}p{0.12\textwidth}|}
\hline
{\small{}$D$} & {\small{}4} & {\small{}5} & {\small{}6} & {\small{}7}\tabularnewline
\hline
{\small{}$\Lambda=0$} & {\small{}0.3361} & {\small{}0.2769} & {\small{}0.2472} & {\small{}0.2287}\tabularnewline
{\small{}$\Lambda=0.001$} & {\small{}0.3393} & {\small{}0.2784} & {\small{}0.2481} & {\small{}0.2293}\tabularnewline
\hline
\end{tabular}
\par\end{centering}{\footnotesize \par}
\end{onehalfspace}

{\footnotesize{}\caption{{\footnotesize{}\label{tab:L}The values of $p_{\star}$ at which
the black hole formations first occur in asymptotically flat spacetimes
and de Sitter spacetimes in different spacetime dimensions.}}
}{\footnotesize \par}
\end{table}
{\small \par}

$p_{\star}$ is the black hole threshold, when $p>p_{\star}$, the dynamical system can end up in the formation of a black hole. From the table above, we can see that in the de Sitter spacetime with cosmological constant, the critical parameter $p_{\star}$ is always bigger than that of the asymptotically flat spacetime. This property keeps the same for different spacetime dimensions. It illustrates that in de Sitter spacetimes, the black hole is more difficult to be formed, which requires the initial scalar perturbation to have stronger amplitude.  This confirms the argument in  \cite{Hod-96} that the positive cosmological constant adds the effect of dispersion. The consistent behavior of the gravitational collapse in de Sitter space in all dimensions also supports the argument of the universal AdS collapse of scalar field in higher dimensions \cite{Jalmuzna-11}.

Besides, from the table we can learn that with the increase of spacetime dimensions, the critical $p_{\star}$ decreases. This keeps for both the asymptotically flat spacetime and de Sitter spacetime. It tells us that with the increase of the spacetime dimensions, the stability property of the original spacetime can be more easily changed and black hole can be more easily developed.  Whether the dependence of the stability of the spacetime on dimensions holds when the cosmological constant is negative is a question to answer.

\subsection{The effect of the charge conjugation $q$}

In this subsection, we focus on the effect of the charge conjugation $q$. To be explicit, we set $D=7$ and then $\Lambda=\frac{(D-1)(D-2)}{2}=15$.
The charge conjugation $q$ describes the coupling strength between the scalar field and the Gauge field. From (18), we see that if we fix the amplitudes of  $\phi_i$ and $\Pi_i(i=1,2)$, the change of the charge conjugation value will modify the conserved electric charge of the system.  In \cite{Hod-96}, the authors discussed this external effect on the gravitational collapse in the asymptotically flat spacetime. They observed that due to the change of the charge conjugation, the black hole charge can always be treated perturbatively at the black hole threshold.

\begin{figure}
\begin{centering}
{\small{}\includegraphics[scale=0.7]{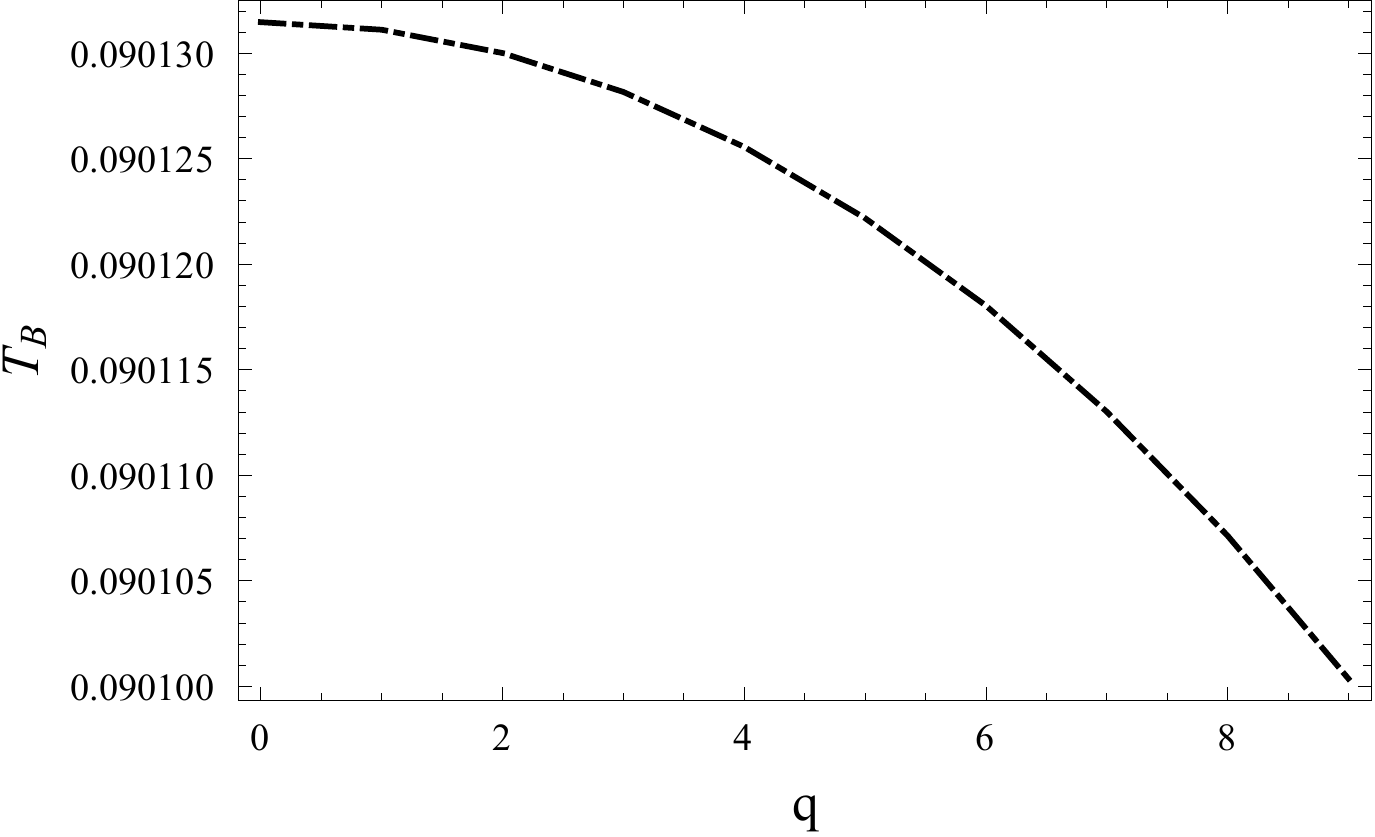}}
\par\end{centering}{\small \par}
{\small{}\caption{{\footnotesize{}\label{fig:TMQ-q}The time needed for the black hole formation for different
$q$, where we fix $a_{1}=0.222$ and $a_{4}=0.3$ in the computation.}}
}{\small \par}
\end{figure}

Here we want to examine their result in the de Sitter spacetime. In doing so, we fix $a_{2},a_{4}$ in the initial conditions (20) of perturbations and we still take $a_3=0$ for simplicity. With the increase of the charge conjugation parameter $q$, we find that the threshold value for $a_1$ to form a black hole becomes smaller. This actually tells us that with the increase of the charge conjugation, the formation of the black hole can be easier. To see more closely, we list the values of the black hole horizon formation time scale, the total electric charge and the total mass in the system with the change of the charge conjugation in Table \ref{tab:q}. In Fig.
\ref{fig:TMQ-q} we plot the relation between the black hole horizon formation time and the charge conjugation $q$ where $a_1$ is chosen over the threshold value for the black hole formation.

We see that when the coupling constant between the scalar field and the gauge field becomes stronger, the electric charge in the system increases.  But
the time $T_{B}$ needed for the formation of black hole decreases.  This explains that the black hole formation becomes easier when the coupling
between the charged scalar field and the electromagnetic field becomes
stronger. This result obtained in the de Sitter space supports the result reported in \cite{Hod-96} found in the asymptotically flat spacetime.
The coupling between the scalar field and the gauge field contributes much more to
the gravitational binding although there is more repulsive effect of the charge as $q$ increases.
In Table \ref{tab:q} we observe the increase of the total mass in the system with the increase of the charge conjugation constant, which is responsible for the more gravitational binding effect.

{\small{}}
\begin{table}
\begin{onehalfspace}
\begin{centering}
{\footnotesize{}}%
\begin{tabular}{|>{\centering}p{0.05\textwidth}|>{\centering}p{0.12\textwidth}|>{\centering}p{0.12\textwidth}|>{\centering}p{0.12\textwidth}|>{\centering}p{0.12\textwidth}|>{\centering}p{0.12\textwidth}|>{\centering}p{0.12\textwidth}|}
\hline
{\footnotesize{}$q$} & {\footnotesize{}0} & {\footnotesize{}2} & {\footnotesize{}4} & {\footnotesize{}6} & {\footnotesize{}8} & {\footnotesize{}10}\tabularnewline
\hline
{\footnotesize{}$Q_{C}$} & {\footnotesize{}0} & {\footnotesize{}$8.01422\times10^{-5}$} & {\footnotesize{}$1.60308\times10^{-4}$} & {\footnotesize{}$2.40479\times10^{-4}$} & {\footnotesize{}$3.20644\times10^{-4}$} & {\footnotesize{}$4.00792\times10^{-4}$}\tabularnewline
{\footnotesize{}$T_{B}$} & {\footnotesize{}0.0901315} & {\footnotesize{}0.09013} & {\footnotesize{}0.0901256} & {\footnotesize{}0.090118} & {\footnotesize{}0.0901071} & {\footnotesize{}0.0900926}\tabularnewline
{\footnotesize{}$M_{C}$} & {\footnotesize{}$4.2811\times10^{-5}$} & {\footnotesize{}$4.2812\times10^{-5}$} & {\footnotesize{}$4.2814\times10^{-5}$} & {\footnotesize{}$4.2818\times10^{-5}$} & {\footnotesize{}$4.2824\times10^{-5}$} & {\footnotesize{}$4.2831\times10^{-5}$}\tabularnewline
\hline
\end{tabular}
\par\end{centering}{\footnotesize \par}
\end{onehalfspace}
{\footnotesize{}\caption{{\footnotesize{}\label{tab:q}The total charge $Q_{C}$, total mass $M_{C}$
and the time scale $T_{B}$ for the formation of a black hole different values of coupling constant $q$. Here we fix $a_{1}=0.222$ and $a_{4}=0.3$ in the computation
 }}
}{\footnotesize \par}
\end{table}
{\small \par}

{\small{}}
{\small \par}

The influence due to the charge conjugation reported above is not specific to the chosen condition in (20). For other forms of the initial conditions used in \cite{Hod-96}, similar observations can also be obtained. 

\subsection{The effect of amplitudes in the initial perturbations }

 We know that the effect of the electric charge opposes gravitation.  In the competition of dynamics the electric charge can play the role of the repulsive force to disperse the perturbation field out. In the charge expression (18), besides the charge conjugation, the initial conditions of perturbations can also determine the value of the conserved charge. Now we fix the charge conjugation value $q=1$ and examine how the electric charge can be influenced by the initial conditions of perturbations and finally in turn affects the gravitational collapse. We set $D=7$ so that  $\Lambda=\frac{(D-1)(D-2)}{2}=15$ in this subsection.

\begin{figure}
\begin{centering}
\includegraphics{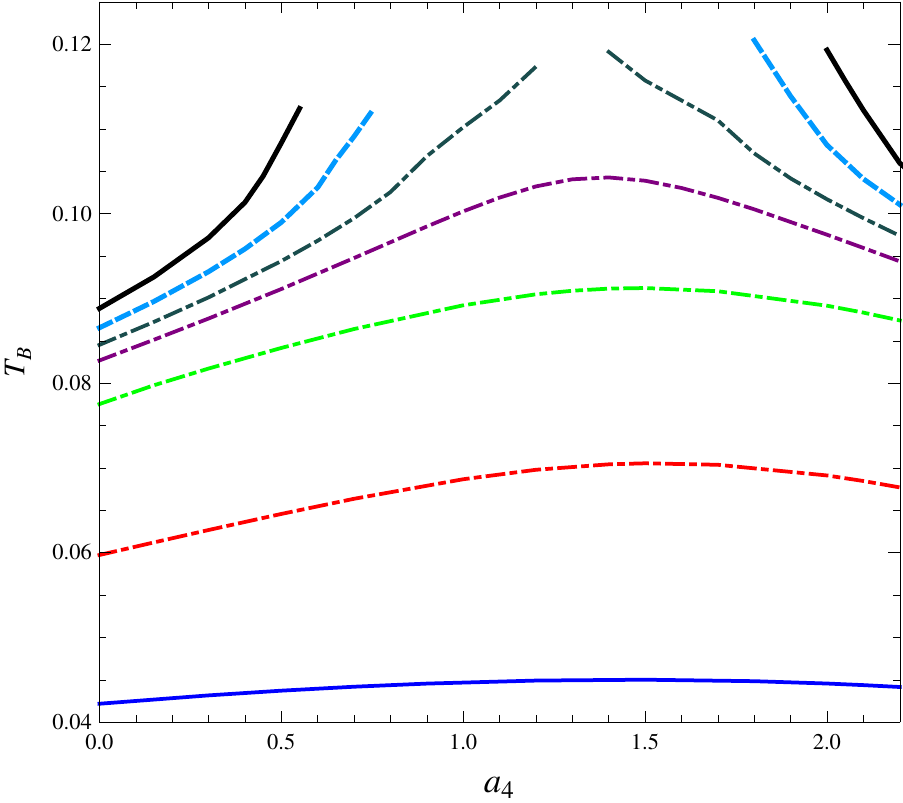}
\par\end{centering}
{\small{}\caption{{\footnotesize{}\label{fig:7D-T-a4} The time needed to form the black hole horizon with the change of $a_4$. The
lines from up to down denote $a_{1}=0.218$, 0.22, 0.222, 0.224, 0.25,
0.3, respectively. }}
}{\small \par}
\end{figure}

Since for simplicity we take $a_3=0$,  from eq.(\ref{eq:Charge-1})
we learn that the total charge value is determined by the product of $a_{1}$ and
$a_{4}$ in our initial profiles. If we fix $a_{1}$, where $a_1$ indicates the amplitude of the initial scalar perturbation, $Q_{C}$ is only
proportional to $a_{4}$. We calculate the influence of $a_4$ (the electric charge value) on the time scale for the black hole horizon to form for different amplitudes of the initial scalar perturbation with different fixed values of $a_{1}$. The numerical results
are shown in Fig. \ref{fig:7D-T-a4}.

It can be seen that the increase of $a_{4}$ leads to the increase of the electric charge and the time needed to develop a black hole increases at first and then starts to decrease. For the scalar perturbation with smaller initial amplitude (smaller $a_{1}$), the different dependence of the time scale for the black hole formation on the increase of $a_4$ parameter is bigger. When the initial amplitude of the perturbation $a_1$ is small enough, for example as shown $a_1<0.224$, the black hole cannot be formed in a certain range of $a_4$. When $a_{1}=0.218$, the black hole cannot be formed in the region $0.57<a_{4}<2$; when $a_{1}=0.22$, the gravitation collapse fails in the interval $0.7<a_{4}<1.8$ and when $a_{1}=0.222$, the black hole cannot develop within  $1.2<a_{4}<1.4$.

This phenomenology is very interesting and has not been found before. We know that the parameter $a_1$ directly relates to the initial amplitude of the perturbation. According to the property of the gravitational collapse disclosed in  \cite{Choptuik-93}, the bigger chosen value of $a_1$ corresponds to the stronger gravitational energy to form a black hole. Here for chosen $a_1$, the value of $a_4$ determines the electric charge. There are two effects with the increase of $a_4$. On the one hand, with the increase of $a_4$, the electric charge increases, the repulsive force due to the electric field starts to play the role and hinders the gravitational collapse. On the other hand, as we can see from Eqs.~(\ref{eq:delta2})(\ref{eq:a2})(\ref{eq:Mass}), the increase of $a_4$ also contributes to the increase of the gravitational energy thus encourages the gravitational collapse. Whether the black hole can be formed or not is determined by the result of the competition between the charge repulsion and gravitational binding.

To examine the physical reason behind the phenomenon more closely, we calculate the gravitational potential since it can be taken as a clear indication to explain whether the black hole can be formed or not. To form a black hole from the gravitational collapse, a very high gravitational barrier is needed to trap the scalar field within a small region; for the scalar field to disperse away, the gravitational potential needs to be  flat. The behaviors of the gravitational potential due to the change of $a_4$  are exhibited in Fig.~4, where we fix $a_1=0.218$ for the discussion. To exhibit clearly the effect of $a_4$ on the gravitational potential, we first find the maximum of the potential in the whole evolution process and mark the corresponding radius $r_f$ and show the evolution of the potential $V(t, r_f)$ at $r=r_f$.

When $a_4<0.57$, we see in the upper left panel of Fig.~4 that the peak of the gravitational potential appears later as $a_4$ increases. This means that, although the black hole can be formed, it needs more bounces of the scalar field to finally collapse and form the black hole with the increase of parameter $a_4$. For $a_{4}=0$, the scalar field needs two bounces while for $a_{4}=0.5$ the scalar field bounces three times. In this case with small value of $a_4$, the dispersion effect due to the electric charge is still minor compared with the gravitational trapping. The behavior of the gravitational potential exactly explains the observed phenomenon shown in Fig.~3 that the time scale for the formation of the black hole becomes longer with the increase of $a_4$ when $a_4$ is small.

Now we look at the upper right panel of Fig.~4, where we concentrate on the parameter $a_4$ within $0.57<a_{4}<1.2$. In this range of $a_4$, the gravitational potential is weak and has no high-enough barrier to trap the perturbation to form the black hole. The dissipation effect due to the electric charge wins the gravitational binding, so that in this range the scalar field finally disperses and escapes out of the cosmological horizon. With the increase of $a_4$ in this range, we see that the gravitational potential becomes even weaker.

In the lower left panel of Fig.~4 we exhibit the potential behavior when the parameter $a_4$ falls in the range  $1.2<a_{4}<1.96$. Although the gravitational potential is still too weak to trap enough perturbation to form a black hole, we observe that with the increase of $a_4$, the peak of the gravitational potential increases. In this parameter range of $a_4$, the repulsive effect still wins the competition of the dynamics so that the black hole cannot be formed.

Now we concentrate on the lower right panel in Fig.~4. The parameter $a_4$ continues to increase to be $a_{4}>2$. We see that a huge barrier of the gravitational potential returns to trap the perturbation. Moreover, with the increase of $a_4$, the peak appears earlier so that fewer bounces are needed for the scalar perturbation to form a black hole. In this case the gravitational potential clearly dominates the competition of dynamics. This actually account for the decrease of the time scale for the formation of a black hole when $a_4$ continues to increase, as shown in Fig.~3.

\begin{figure}
\begin{centering}
{\small{}}%
\begin{tabular}{cc}
{\small{}\includegraphics[scale=0.8]{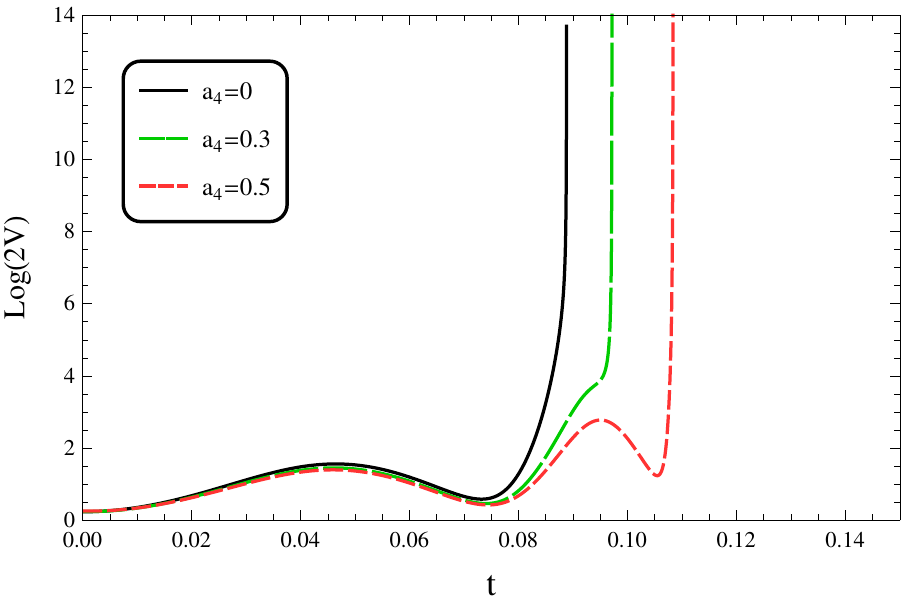}} & \includegraphics[scale=0.8]{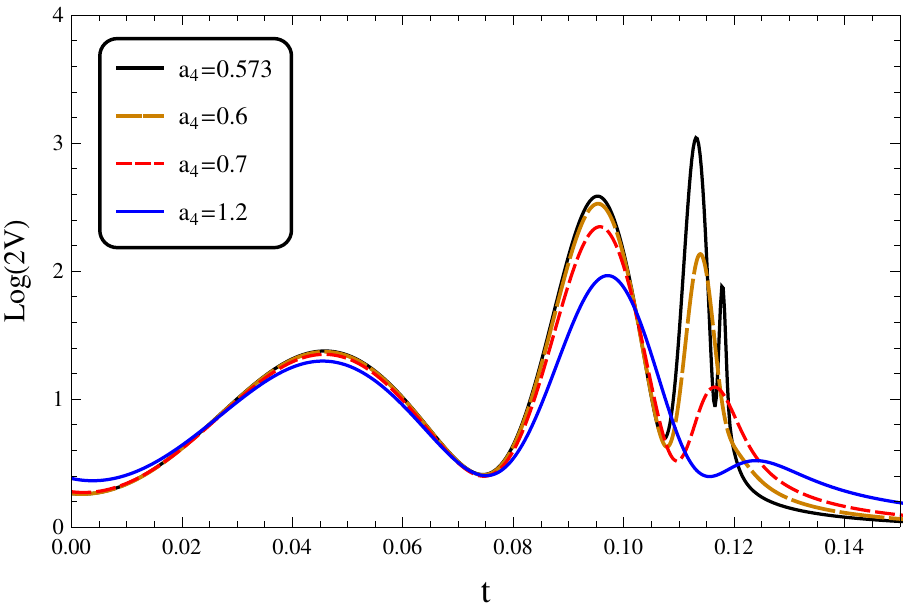}\tabularnewline
{\small{}\includegraphics[scale=0.8]{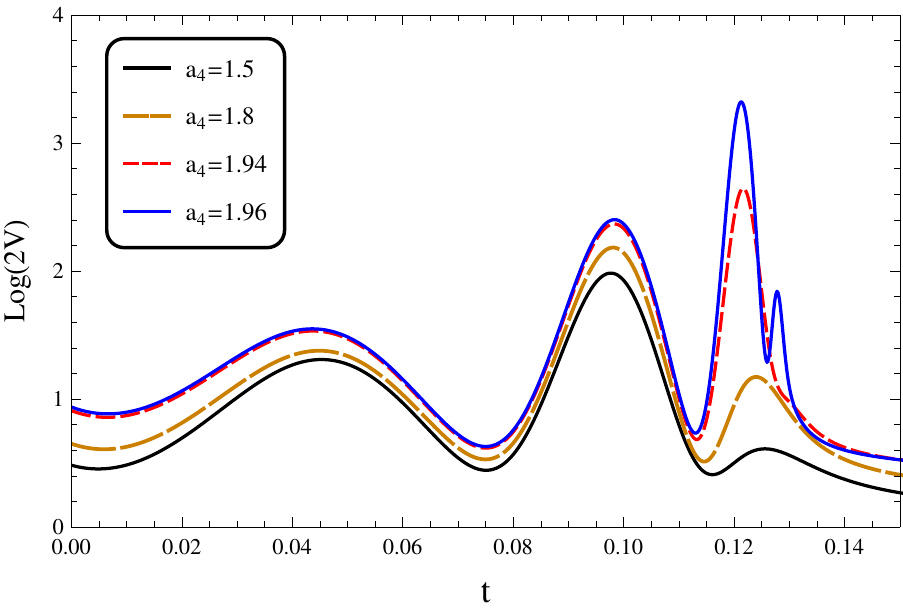}} & \includegraphics[scale=0.8]{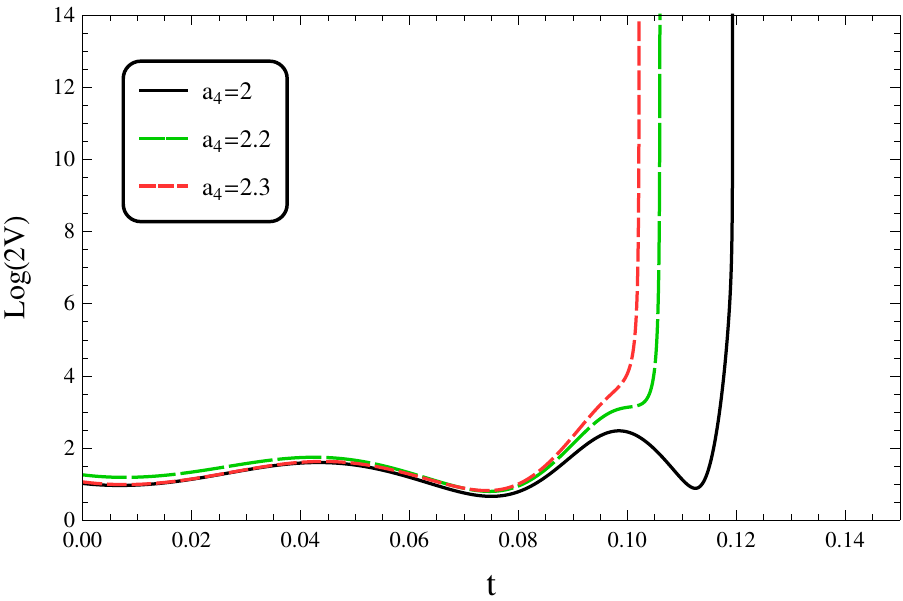}\tabularnewline
\end{tabular}
\par\end{centering}{\small \par}

{\small{}\caption{{\footnotesize{}\label{fig:potential_2} In the left column we show the time evolution of the gravitational potential behavior when we fix
$a_{1}=0.218$. In the right column we plot the potential of the gauge field.}}
}{\small \par}
\end{figure}

It is interesting to show that the phenomenon we discovered is not specific to the initial conditions (\ref{eq:initial2}), we have also checked other initial conditions, for example
\begin{eqnarray}
\phi_{1}(0,r)=a_{1}e^{-\left(\frac{r}{\sigma_{1}}\right)^{2}}\cos(r) & , & \phi_{2}(0,r)=a_{2}e^{-\left(\frac{r}{\sigma_{2}}\right)^{2}}\cos(r),\nonumber \\
\Pi_{1}(0,r)=a_{3}e^{-\left(\frac{r}{\sigma_{3}}\right)^{2}}\cos(r) & , & \Pi_{2}(0,r)=a_{4}e^{-\left(\frac{r}{\sigma_{4}}\right)^{2}}\cos(r),\label{eq:initial3}
\end{eqnarray}
and also for other forms of spherically symmetric initial conditions.
We find the similar behavior as shown in Figure 3. So the phenomenon we discovered above is not specific to the initial conditions (\ref{eq:initial2}). It holds for a rather general class of initial perturbations.

\subsection{Other dimensions}
The above results are for the seven-dimensional de Sitter spacetime. Considering that in the study of the linear perturbation of the highly charged de Sitter black hole, it was found that the instability appears when the spacetime dimension is bigger than seven  while disappears for lower dimensional configurations \cite{Konoplya-08}, we have the motivation to  check whether the gravitational collapse of the charged scalar field in de Sitter spacetime can also exhibit special property in different dimensions.

For this reason,  we have generalized the above calculations to other dimensions from $D=4$ to $D=9$ with cosmological constant in the form $\Lambda=\frac{(D-1)(D-2)}{2}$.  We found from numerical computations,  the qualitative behaviors of the collapse of the charged scalar
field in other dimensions are similar to those reported above in seven dimensions.

The universality of the gravitational collapse in charged scalar field in different dimensional de Sitter spacetimes is consistent with the study of the collapse of neutral scalar field in the same de Sitter space-time. Combining with the results obtained here and in  \cite{Jalmuzna-11,Garfinkle-99}, we can conclude that the dimension of the space-time cannot change the qualitative property in the nonlinear results in the gravitational collapse.

\section{Conclusions}

We have studied the charged scalar collapse in the spherically symmetric de Sitter spacetimes. The de Sitter space-time with positive cosmological constant is different from the asymptotically flat space-time since the positive cosmological constant
contributes a dissipation effect against the imploding of the perturbation.  This is the reason that with the cosmological constant, the process of the gravitational collapse is hindered and it requires bigger critical threshold for the formation of a black hole. We have shown that this property is universal in higher dimensions.

Besides the cosmological constant, we have also examined another external factor, the electric charge, to influence the gravitational collapse.  Dynamically with the addition of the electric charge, there are three competitors competing in the process of the gravitational collapse:
gravity traps the scalar field, kinetic energy and electromagnetic force disperse the scalar field. The electric charge effect on the gravitational collapse can be affected by the charge conjugation strength which was first discovered in the asymptotically flat space-time in \cite{Hod-96}. Here we have confirmed their result in the de Sitter spacetime that the electric charge effect due to the change of the charge conjugation is perturbative in the threshold of the black hole formation. Furthermore, we have disclosed a new result that the electric charge effect in the dynamical competition can be completely controlled by the initial conditions in perturbations. This is similar to the other two effects caused by the gravitational energy and kinetic energy, respectivily. Tuning the parameters in the initial conditions of the charged scalar perturbation, we can obtain richer physics. Instead of always being the perturbative effect for the electric charge due to the change of the charge conjugation, the increase of the electric charge results from the change of the initial conditions can present us the  repulsive effect to prevent the gravitational collapse.  But when the gravitational binding energy dominates, the further increase of the electric charge due to the change of the initial condition of perturbation cannot change the property of the formation of a black hole.

We have checked the properties of stability in the gravitational collapse for the charged scalar field perturbation in spherically symmetric de Sitter spacetime in different dimensions and found that    the phenomenon we discovered holds for a rather general class of initial perturbations.  Combining with the universal results for the gravitational collapse in asymptotically flat spacetime  \cite{Garfinkle-99} and anti-de Sitter spacetime   \cite{Jalmuzna-11} in various dimensional spacetimes, we can conclude that the dynamics of the gravitational collapse is qualitatively  independent  of spacetime dimension.  This is different from the stability property of the highly charged Reissner-Nordstrom black hole in de Sitter spacetimes.  At this moment we cannot give the reason on the differences. The stability of highly charged black hole in de Sitter spacetime was obtained from the linear perturbation study, while the gravitational collapse result was derived from the nonlinear perturbation theory. We do not know whether the nonlinear perturbation of the stability analysis can kill the dimensional influence on  the property in the stability of the  highly charged Reissner-Nordstrom-de Sitter black holes. To answer this question, more computations are needed.

\section*{Acknowledgments}

We thank Eleftherios Papantonopoulos, Li-Wei Ji and Yun-Qi Liu for helpful discussions. This work
is supported by NNSF of China.


\begin{thebibliography}{10}
\bibitem{Choptuik-93}M. W. Choptuik,\textit{ Universality and scaling
in gravitational collapse of a massless scalar field}, Phys. Rev.
Lett. 70, 9 (1993).

\bibitem{AnZhong-01}A. Wang, \textit{Critical phenomena in gravitational collapse: The Studies so far}, Braz.J.Phys. 31 (2001) 188-197,
{[}{[}arXiv:gr-qc/0104073{]}.

\bibitem{Gundlach-07}C. Gundlach, J. M. Martin-Garcia, \textit{Critical
phenomena in gravitational collapse}, Living Rev.Rel. 10 (2007) 5,
{[}{[}arXiv:0711.4620 {[}gr-qc{]}{]}.

\bibitem{Choptuik-98}M.W. Choptuik, \textit{The (Unstable) threshold of black hole formation}, Talk given at Conference: C97-12-16,
{[}arXiv:gr-qc/9803075{]}.

\bibitem{Bizon-11}P. Bizon, A. Rostworowski, \textit{On weakly turbulent
instability of anti-de Sitter space}, Phys.Rev.Lett. 107 (2011) 031102,
{[}arXiv:1104.3702 {[}gr-qc{]}{]}.

\bibitem{Jalmuzna-11}J. Jalmuzna, A. Rostworowski, P. Bizon, \textit{A
Comment on AdS collapse of a scalar field in higher dimensions}, Phys.Rev.
D84 (2011) 085021, {[}arXiv:1108.4539 {[}gr-qc{]}{]}.

\bibitem{Garfinkle-11}D. Garfinkle, L. A. Pando Zayas , \textit{Rapid Thermalization in Field Theory from Gravitational Collapse}, Phys.Rev. D84 (2011) 066006, {[}arXiv:1106.2339 {[}hep-th{]}{]}.

\bibitem{Deppe-14}N. Deppe, A. Kolly, A. Frey, G. Kunstatter,\textit{
Stability of AdS in Einstein Gauss Bonnet Gravity}, Phys.Rev.Lett.
114 (2015) 071102, {[}arXiv:1410.1869 {[}hep-th{]}{]}.

\bibitem{Cai-15}R.-G. Cai, L.-W. Ji, R.-Q. Yang, \textit{Collapse
of self-interacting scalar field in anti-de Sitter space}, {[}arXiv:1511.00868
{[}gr-qc{]}{]}.

\bibitem{Hod-96}S. Hod, T. Piran, \textit{Critical behavior and universality
in gravitational collapse of a charged scalar field}, Phys.Rev. D55
(1997) 3485-3496, {[}arXiv: gr-qc/9606093{]}.

\bibitem{Iwashita-05}Y. Iwashita, H. Yoshino, T. Shiromizu, \textit{Gravitational
collapse disturbs the dS/CFT correspondence?} Phys.Rev. D72 (2005)
084014, {[}gr-qc/0507076{]}.

\bibitem{Gundlach-96}C. Gundlach, J. M. Martin-Garcia, \textit{Charge
scaling and universality in critical collapse}, Phys.Rev. D54 (1996)
7353-7360, {[}arXiv: gr-qc/9606072{]}.

\bibitem{Nakonieczna-11}A. Nakonieczna, M. Rogatko, R. Moderski, \textit{Collapse of Charged Scalar Field in Dilaton Gravity}, Phys.Rev. D83 (2011) 084007, {[}arXiv:1103.4808
{[}hep-th{]}{]}.

\bibitem{Nakonieczna-12}A. Nakonieczna, M. Rogatko, R. Moderski, \textit{Dynamical Collapse of Charged Scalar Field in Phantom Gravity},  Phys.Rev. D86 (2012) 044043 , {[}arXiv:1209.1203
{[}hep-th{]}{]}.

\bibitem{Nakonieczna-15}A. Nakonieczna, M. Rogatko, L. Nakonieczny, \textit{Dark sector impact on gravitational collapse of an electrically charged scalar field}, JHEP11(2015)012, {[}arXiv:1508.02657
{[}hep-th{]}{]}.


\bibitem{Konoplya-08} R.A. Konoplya, A. Zhidenko,\textit{ Instability
of higher dimensional charged black holes in the de-Sitter world},
Phys.Rev.Lett. 103 (2009) 161101, {[}arXiv:0809.2822 {[}hep-th{]}{]}.

\bibitem{Cardoso-10}V. Cardoso, M. Lemos, M. Marques,\textit{ On
the instability of Reissner-Nordstrom black holes in de Sitter backgrounds},
Phys.Rev. D80 (2009) 127502, {[}arXiv:1001.0019 {[}gr-qc{]}{]}.

\bibitem{Garfinkle-99}D. Garfinkle, C. Cutler, G.C. Duncan, \textit{Choptuik scaling in six-dimensions}, Phys.Rev. D60 (1999) 104007, {[}arXiv:gr-qc/9908044{]}.

\bibitem{Maliborski-13} M. Maliborski and A. Rostworowski, \textit{Lecture
Notes on Turbulent Instability of Anti-de Sitter Spacetime}, Int.
J. Mod. Phys. A \textbf{28}, 1340020 (2013), {[}arXiv:1308.1235 {[}gr-qc{]}{]}.

\bibitem{Kreiss-72}H.-O. Kreiss and J. Oliger, \textit{Comparison
of accurate methods for the integration of hyperbolic equations},
Tellus 24 (1972) no. 3, 199\textendash 215.

\end{thebibliography}
\end{document}